\documentclass[12pt]{iopart}
\usepackage{graphicx}
\usepackage{esdiff}
\usepackage{subfig}
\usepackage{comment}
\usepackage{hhline}
\usepackage{array}
\usepackage{textcomp}
\usepackage{hyphsubst}
\usepackage{hyperref}

\begin{document}

\title[Analytical model of the readout power dependence of microwave SQUID multiplexers]{Analytical model of the readout power and SQUID hysteresis parameter dependence of the resonator characteristics of microwave SQUID multiplexers}

\author{M Wegner$^{1,2,3}$, C Enss$^{1,3}$ and S Kempf$^{1,2}$}
\address{$^1$ Kirchhoff-Institute for Physics, Heidelberg University, Im Neuenheimer Feld 227, D-69120 Heidelberg, Germany}
\address{$^2$ Institute of Micro- and Nanoelectronic Systems, Karlsruhe Institute of Technology, Hertzstrasse 16, Building 06.41, D-76187 Karlsruhe, Germany}
\address{$^3$ Institute for Data Processing and Electronics, Karlsruhe Institute of Technology, Hermann-von-Helmholtz-Platz 1, Building 242, D-76344 Eggenstein-Leopoldshafen}
\ead{\mailto{mathias.wegner@kit.edu}}
\vspace{10pt}
\begin{indented}
\item[]December 2021
\end{indented}

\begin{abstract}
We report on the development of an analytical model describing the readout power and SQUID hysteresis parameter dependence of the resonator characteristics used for frequency encoding in microwave SQUID multiplexers. Within the context of this model, we derived the different dependencies by analyzing the Fourier components of the non-linear response of the non-hysteretic rf-SQUID. We show that our model contains the existing model as a limiting case, leading to identical analytical expressions for small readout powers. Considering the approximations we made, our model is valid for rf-SQUID hysteresis parameters $\beta_{\mathrm{L}} < 0.6$ which fully covers the parameter range of existing multiplexer devices. We conclude our work with an experimental verification of the model. In particular, we demonstrate a very good agreement between measured multiplexer characteristics and predictions based on our model.
\end{abstract}

\vspace{2pc}

\maketitle

\section{Introduction}
\label{sec:1}

Superconducting quantum interference devices (SQUIDs) are the devices of choice for measuring any physical quantity that can be naturally converted into a magnetic flux change. They provide noise levels close to the quantum limit as well as ultra-low power dissipation down to a few pW for single devices \cite{Dru07, Cla04}. SQUIDs are hence intrinsically compatible with sub-Kelvin operation temperatures and are thus particularly suited for reading out cryogenic particle detectors such as transition edge sensors (TESs) \cite{Ull15} or magnetic microcalorimeters (MMCs) \cite{Kem18}. Both, TESs and MMCs, are calorimetric single-particle detectors converting an energy input upon the absorption of an energetic particle into a temperature rise that is continuously monitored via an ultra-sensitive thermometer that is based either on the steep temperature dependence of the resistance within the S/N transition of a superconducting material (TESs) or the temperature dependence of a paramagnetic material that is situated in a weak external bias magnetic field. The unique combination of such ultra-sensitive thermometers and a SQUID based readout chain ultimately yields energy-dispersive single-particle detectors with outstanding energy resolution, fast signal rise time, a quantum efficiency close to 100\% and a large dynamic range.

The maturity of fabrication technologies allows building large-scale detector systems employing thousands or even millions of virtually identical detectors paving the way for realizing next-generation instruments such as the Simons observatory ~\cite{Ade19} or experiments investigating the electron neutrino mass \cite{Alp15, Gas17} with sub-eV/$c^2$ mass sensitivity. However, for this kind of instruments, system complexity, cooling power at the cryogenic platform as well as overall costs has to be taken into account. For these reasons, SQUID based multiplexing techniques turn out to be key technologies for these applications.

Existing SQUID-based cryogenic multiplexers rely on time-division multiplexing (TDM) \cite{Dor16}, frequency-division multiplexing (FDM) using MHz \cite{Har14} and GHz \cite{Mat08} carriers, code-division multiplexing (CDM) \cite{Mor16} using orthogonal Walsh codes or hybrid techniques \cite{Rei08}. Among those techniques, microwave SQUID multiplexing ($\mu$MUXing)~\cite{Irw04} is thought to be best suited for realizing ultra-large scale detector arrays since this technique offers a large multiplexing factor, the required bandwidth per readout channel for fast calorimetric detectors, a very good noise performance as well as an extremely low on-chip power dissipation. However, realizing and in particular optimizing such complex readout systems requires detailed models describing the device physics to reliably predict the performance of the readout system on the basis of design and operation parameters.

Existing models lead to an accurate description of the $\mu$MUX characteristics for small readout powers, i.e. $P_\mathrm{rf} \rightarrow 0$ \cite{mat11, kem17}. For experimentally more realistic values, i.e. when using a $\mu$MUX for cryogenic detector readout, large discrepancies between measurements and model predictions are observed \cite{kem17}. For this reason, existing $\mu$MUX models can't reliably be used for optimizing the device performance. However, operating the cryogenic multiplexer with optimal parameters is of utmost importance as its performance influences the overall system noise level and potentially might limit the achievable energy resolution \cite{Kem18}.

Against this background, we present a $\mu$MUX model that precisely allows predicting the characteristics of a $\mu$MUX for a wide parameter range of the readout signal amplitude and the SQUID hysteresis parameter. We compare our model with existing models reliably predicting the  $\mu$MUX performance for small readout powers to determine parameter constraints and show that our model includes existing models as a limiting case. We conclude our work with an experimental verification of our model. Here, we demonstrate an excellent agreement between predictions based on our model and measured $\mu$MUX characteristic curves.

\section{Basics of microwave SQUID multiplexing}
\label{sec:2}

\begin{figure}[t]
    \centering
    \includegraphics[width=8cm]{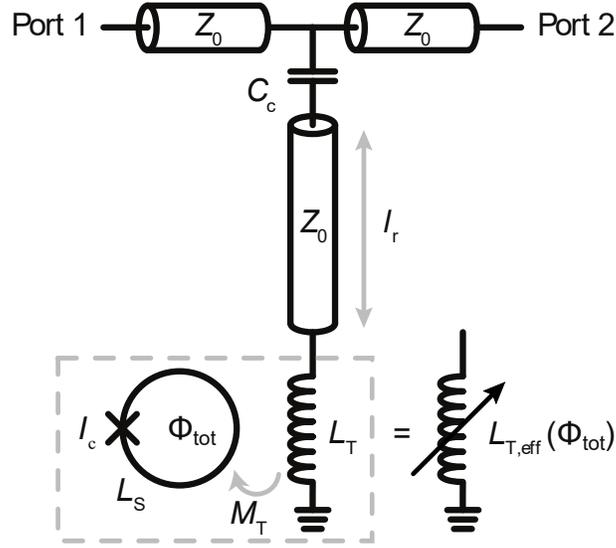}
    \caption{Schematic circuit diagram of a single $\mu$MUX readout channel. The two main components are (i) a non-hysteretic rf-SQUID consisting of a closed superconducting loop with loop inductance $L_{\mathrm{S}}$ and a Josephson tunnel junction with critical current $I_{\mathrm{c}}$ and (ii) a coplanar, quarter-wave resonator with geometrical length $l_{\mathrm{r}}$ that is coupled to a microwave transmission line via a coupling capacitor $C_{\mathrm{c}}$ and shorted to ground via a load inductor $L_{\mathrm{T}}$. The effective load inductance $L_{\mathrm{T,eff}}$ takes into account the magnetic field dependence of the load inductor $L_\mathrm{T}$ due to the mutual interaction with the rf-SQUID.}
    \label{fig:1}
\end{figure}

Figure~\ref{fig:1} shows the schematic circuit diagram of a single $\mu$MUX readout channel. It consists of a coplanar, quarter-wave resonator with characteristic impedance $Z_{0}$ and physical length $l_{\mathrm{r}}$. The resonance frequency of the unloaded resonator is
\begin{equation}
f_{0} = \frac{1}{4l_{\mathrm{r}}\sqrt{\left(L^{'}_{\mathrm{m}}+L^{'}_{\mathrm{kin}}\right)C^{'}}},
\label{eq:1}
\end{equation}
where $L^{'}_{\mathrm{m}}$, $L^{'}_{\mathrm{kin}}$ and $C^{'}$ denote the geometrical (magnetic) inductance, the kinetic inductance of the Cooper pairs and the geometrical (electric) capacitance per unit length, respectively. The open end of the resonator is coupled to a microwave transmission line with characteristic impedance $Z_{0}$ via a coupling capacitor with capacitance $C_{\mathrm{c}}$. The other end of the resonator is shorted to ground via a load inductor with inductance $L_{\mathrm{T}}$ that is simultaneously used to weakly couple a non-hysteretic rf-SQUID to the resonator. The rf-SQUID consists of a superconducting loop with inductance $L_{\mathrm{S}}$ which is interrupted by a single Josephson tunnel junction with critical current $I_{\mathrm{c}}$. To guarantee for a non-hysteretic behavior, the SQUID hysteresis parameter $\beta_{\mathrm{L}} = 2 \pi L_{\mathrm{S}} I_{\mathrm{c}} / \Phi_{0}$ is set to  $\beta_{\mathrm{L}} \le 1$. Here, $\Phi_{0} = 2.07\cdot10^{-15}\,\mathrm{Vs}$ denotes the magnetic flux quantum. The coupling strength between resonator and rf-SQUID is quantified by the mutual inductance $M_{\mathrm{T}}$.

For a vanishing coupling strength between resonator and rf-SQUID, i.e. $M_{\mathrm{T}} \rightarrow 0$, the input impedance of the loaded resonator is given by the series connection of the coupling capacitance and the terminated resonator:
\begin{equation}
Z_{\mathrm{in}} = \frac{1}{i \omega C_{\mathrm{c}}}+ Z_{\mathrm{0}}\frac{i \omega L_{\mathrm{T}}+Z_{\mathrm{0}}\tanh\left(\gamma l_{\mathrm{r}}\right)}{Z_{0}+i \omega L_{\mathrm{T}}\tanh\left(\gamma l_{\mathrm{r}}\right)}.
\label{eq:2}
\end{equation}
Here, $\gamma$ describes the complex propagation constant of the electromagnetic wave with angular frequency $\omega$ inside the resonator. Using the condition $\mathrm{Im}(Z_{\mathrm{in}}) = 0$, we yield the expression  
\begin{equation}
    f_{\mathrm{r}} \approx \frac{f_{0}}{1+4 f_{0}\left(C_{\mathrm{c}}Z_{0}+L_{\mathrm{T}}/Z_{0}\right)}
    \label{eq:3}
\end{equation}
for the resonance frequency of the loaded resonator which is shifted towards smaller frequencies as compared to the resonance frequency $f_{0}$ of the unloaded resonator.

We describe the mutual interaction between the rf-SQUID and the resonator by introducing an effective impedance of the resonator termination. Here, for simplicity, we model the Josephson junction as a pure non-linear inductor with inductance $L_{\mathrm{JJ}}(\varphi_{\mathrm{tot}}) =  \Phi_{0}/\left[2\pi I_{\mathrm{c}} \cos\left(\varphi_{\mathrm{tot}}\right)\right]$, which depends on the normalized total magnetic flux $\varphi_{\mathrm{tot}}=2\pi\Phi_{\mathrm{tot}}/\Phi_{0}$ threading the SQUID loop, and hence neglect the subgap resistance $R_\mathrm{sg}$ and the intrinsic capacitance $C_\mathrm{JJ}$, as their contribution is about an order of magnitude smaller for most devices. In this case, the inductance $L_{\mathrm{T,eff}}$ of the effective load inductor is given by the expression
\begin{equation}
L_{\mathrm{T,eff}}(\varphi_{\mathrm{tot}}) =  L_{\mathrm{T}} - \frac{M_{\mathrm{T}}^2}{L_{\mathrm{S}}+L_{\mathrm{JJ}}(\varphi_{\mathrm{tot}})} =  L_{\mathrm{T}} - \frac{M_{\mathrm{T}}^2}{L_{\mathrm{S}}} \frac{\beta_{\mathrm{L}} \cos\left(\varphi_{\mathrm{tot}}\right)}{1 + \beta_{\mathrm{L}} \cos\left(\varphi_{\mathrm{tot}} \right)}.
\label{eq:4}
\end{equation}
Using equation~(\ref{eq:3}) and replacing $L_{\mathrm{T}}$ by $L_{\mathrm{T,eff}}(\varphi_{\mathrm{tot}})$ allows deriving an approximation for the magnetic flux dependence of the resonance frequency of the loaded resonator:
\begin{equation}
f_{\mathrm{r}}(\varphi_{\mathrm{tot}}) 
\approx f_{0}-4 f_{0}^2\left[C_{\mathrm{c}}Z_{0}+ \frac{L_{\mathrm{T}}}{Z_{0}}- \frac{M_{\mathrm{T}}^2}{Z_{0} L_{\mathrm{S}}} \frac{\beta_{\mathrm{L}} \cos\left(\varphi_{\mathrm{tot}}\right)}{1 + \beta_{\mathrm{L}} \cos\left(\varphi_{\mathrm{tot}}\right)}\right].
\label{eq:5}
\end{equation}
Here, we assumed weak capacitive coupling, i.e. $\omega_{0} C_{\mathrm{c}} Z_{0} \ll 1$ with $\omega_{0} = 2\pi f_{0}$, and small frequency shifts caused by the SQUID, i.e. $\omega_{0} L_{\mathrm{T}} \ll Z_{0}$.

\begin{figure}[t]
    \centering
    \includegraphics[width=9cm]{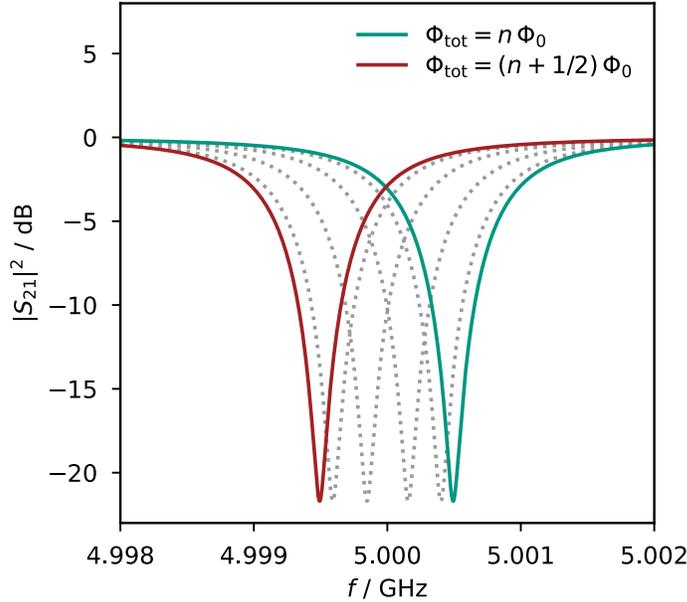}
    \caption{Frequency dependence of the transmission amplitude $|S_{21}(f)|^{2}$ of a single multiplexer channel for different values of the magnetic flux $\Phi_{\mathrm{tot}}$ threading the SQUID loop. The colored curves indicate the curves with highest (green line) and lowest (red line) resonance
    frequency. The parameters used for calculating the curves are $f_{\mathrm{r}} = 5\,\mathrm{GHz}$, $\Delta f_{\mathrm{BW}} = 1\,\mathrm{MHz}$, $Z_{0} = 50\,\Omega$, $M_{\mathrm{T}}^{2}/L_{\mathrm{S}} = 25.4\,\mathrm{pH}$ and $\beta_{\mathrm{L}}=0.01$, leading to a maximum resonance frequency shift of $\Delta f_{\mathrm{r}}^{\mathrm{max}} = 1\,\mathrm{MHz}$. Here, $\Delta f_\mathrm{BW}$ describes the resonator bandwidth. The total magnetic flux~$\Phi_{\mathrm{tot}}$ is varied between $n\Phi_0$ and $(n+1/2)\Phi_0$ in steps of $0.1\,\Phi_{0}$.}
    \label{fig:2}
\end{figure}

To illustrate the flux dependence of the resonance frequency, figure \ref{fig:2} shows the transmission amplitude $|S_{21}(f)|^{2} = |2/\left(2+Z_{0}/Z_{\mathrm{in}}(f)\right)|^{2}$ between port 1 and port 2 of an exemplary resonator for different values of the magnetic flux $\Phi_{\mathrm{tot}}$. As one can see, the resonance frequency oscillates in between its maximum for $\Phi_{\mathrm{tot}} = n\,\Phi_{0}$ and its minimum for $\Phi_{\mathrm{tot}} = (n+1/2)\,\Phi_{0}$. For a magnetic flux of $\Phi_{\mathrm{tot}} = (n\pm 1/4)\,\Phi_{0}$, the Josephson inductance $L_{\mathrm{JJ}}(\varphi_{\mathrm{tot}})$ diverges, and therefore the resonance frequency $f_{\mathrm{r}}([n\pm 1/4]\,\Phi_{0}) = 5\,\mathrm{GHz}$ is simply given by equation (\ref{eq:3}) since the resonance frequency is not affected by means of the rf-SQUID. 

\begin{figure}[t]
    \centering
    \includegraphics[width=9cm]{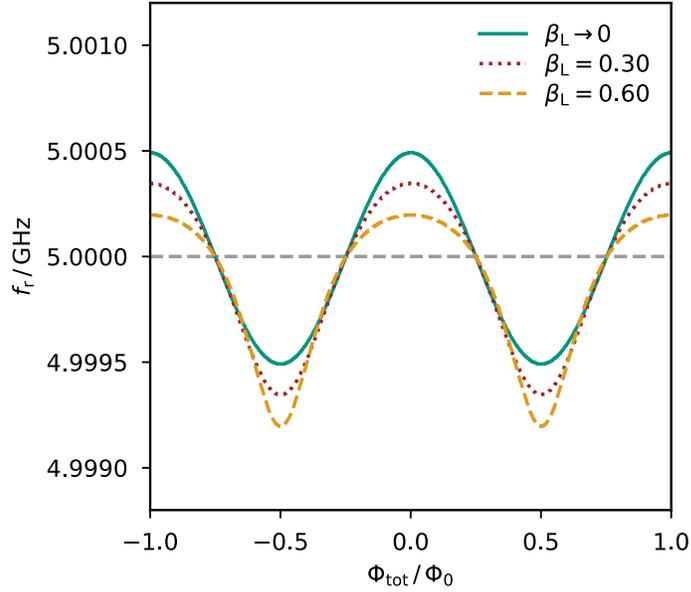}
    \caption{Flux dependence of the resonance frequency $f_{\mathrm{r}}(\Phi_{\mathrm{tot}})$ for different values of the SQUID hysteresis parameter $\beta_{\mathrm{L}}$. The ratio $M_{\mathrm{T}}^{2}/L_{\mathrm{S}}$ is adapted to the corresponding value of $\beta_{\mathrm{L}}$ in such a way that the maximum resonance frequency shift $\Delta f_{\mathrm{r}}^{\mathrm{max}} = 1\,\mathrm{MHz}$ stays constant. All other parameters of the curves correspond to the values chosen in figure \ref{fig:2}.}
    \label{fig:3}
\end{figure}

In figure \ref{fig:3} the impact of the hysteresis parameter $\beta_{\mathrm{L}}$ on the flux dependent resonance frequency shift $f_{\mathrm{r}}(\Phi_{\mathrm{tot}})$ is shown. For very small values of the SQUID hysteresis parameter, i.e. $\beta_{\mathrm{L}} \to 0$, screening currents and therefore self-induced magnetic flux contributions $\Phi_{\mathrm{scr}}$ within the SQUID loop are negligibly small. The characteristics are hence highly symmetric and describe an ideal sinusoidal behavior around its center frequency. For larger values of $\beta_{\mathrm{L}}$, screening currents within the SQUID loop become relevant leading to an asymmetric response and to a general shift of the resonance frequency towards smaller frequencies. 

\section{Revised model for describing the characteristics of a microwave SQUID multiplexer}
\label{sec:3}

The existing $\mu$MUX model can be used to accurately determine the $\mu$MUX characteristics in the limit of very low readout powers, i.e. $P_\mathrm{rf}\rightarrow 0$. However, large deviations between measured characteristics and model predictions are observed for increasingly high readout powers \cite{kem17}. The $\mu$MUX model described in the following is able to describe the effects occurring at high readout powers, thus allowing to describe the $\mu$MUX characteristics over the entire practically used readout power range.

\subsection{Effective load inductance}
\label{sec:3.1}

An electromagnetic wave with power~$P_{\mathrm{rf}}$ and angular frequency $\omega$ being close to the resonance frequency of the resonator, i.e. $\omega \approx 2 \pi f_{\mathrm{r}}$, and traversing from port 1 to port 2 of the microwave transmission line leads to an oscillating current~$i_{\mathrm{T}}(t) = I_{\mathrm{T}} \sin(\omega t)$ within the inductance~$L_{\mathrm{T}}$ terminating the microwave resonator. On resonance, i.e. for $\omega = 2 \pi f_{\mathrm{r}}$, the amplitude of the current reaches the maximum value
\begin{equation}
I_{\mathrm{T}} = \sqrt{\frac{8}{\pi} \frac{ Q_{\mathrm{l}}^{2}}{Q_{\mathrm{c}}}\frac{P_{\mathrm{rf}}}{  Z_{0}}}.
\label{eq:6}
\end{equation}
Here, $Q_{\mathrm{l}}$ and $Q_{\mathrm{c}}$ denote the loaded quality factor and the coupling quality factor of the resonator, respectively. Due to the mutual inductance~$M_{\mathrm{T}}$ between resonator termination and SQUID loop, the current~$i_{\mathrm{T}}(t)$ generates a sinusoidal magnetic flux signal with amplitude $\Phi_{\mathrm{rf}} = M_{\mathrm{T}} I_{\mathrm{T}}$ in the rf-SQUID. Considering an additional, external quasi-static magnetic flux contribution~$\Phi_{\mathrm{ext}}$ as induced by a signal source to be measured, e.g. caused by an inductively coupled input coil, the current in the rf-SQUID can be expressed as
\begin{equation}
I_{\mathrm{S}}(t) = - I_{\mathrm{c}} \sin\left[\varphi_{\mathrm{ext}} + \varphi_{\mathrm{rf}}\sin\left(\omega t\right) + \beta_{\mathrm{L}} \frac{I_{\mathrm{S}}(t)}{I_{\mathrm{c}}} \right].
\label{eq:7}
\end{equation}
Here, $\varphi_{\mathrm{ext}} = 2\pi \Phi_{\mathrm{ext}}/\Phi_{0}$ and $\varphi_{\mathrm{rf}} = 2\pi \Phi_{\mathrm{rf}}/\Phi_{0}$ denote normalized magnetic flux values. The last term in equation~(\ref{eq:7}) is the normalized magnetic flux~$\varphi_{\mathrm{scr}}(t) = 2 \pi I_{\mathrm{S}}(t) L_{\mathrm{S}} / \Phi_{0} = \beta_{\mathrm{L}} I_{\mathrm{S}}(t) / I_{\mathrm{c}}$ that is induced by the supercurrent running in the SQUID loop. The supercurrent $I_{\mathrm{S}}(t)$ inductively couples to the resonator termination, creating a high frequency flux signal~$\Phi_{\mathrm{T}}(t) = M_{\mathrm{T}} I_{\mathrm{S}}(t)$ within the termination. Therefore, in accordance to Lenz's law, the voltage~$u_{\mathrm{ind}} (t) = - M_{\mathrm{T}}\,\mathrm{d} I_{\mathrm{S}}(t) / \mathrm{d} t$ and hence the current 
\begin{equation}
i_{\mathrm{ind}}(t) = -\frac{M_{\mathrm{T}}}{i\omega L_{\mathrm{T}}} \diff{ I_{\mathrm{S}}(t)} {t}
\label{eq:8}
\end{equation}
are induced in the resonator termination. The total current~$i_{\mathrm{tot}} (t) = i_{\mathrm{T}}(t) + i_{\mathrm{ind}}(t)$ in the resonator termination is hence a superposition of two contributions originating from the microwave signal probing the resonator and the supercurrent flowing within the SQUID loop. The voltage across the resonator termination, $u_{\mathrm{tot}}(t) = L_{\mathrm{T}} \, \mathrm{d}i_{\mathrm{tot}}(t) / \mathrm{d}t$, can hence be expressed by introducing an effective inductance
\begin{equation}
L_{\mathrm{T,eff}} = L_{\mathrm{T}} \frac{i_{\mathrm{tot}}(t)}{i_{\mathrm{T}}(t)} = L_{\mathrm{T}}\left(1+\frac{i_{\mathrm{ind}}(t)}{i_{\mathrm{T}}(t)}\right)
\label{eq:9}
\end{equation}
of the resonator termination.

\subsection{\texorpdfstring{$\mu$}{µ}MUX characteristics for small readout powers\texorpdfstring{, i.e. $P_\mathrm{rf}\rightarrow 0$}{}}
\label{sec:3.2}

The characteristics of a $\mu$MUX for small readout powers, i.e. $\varphi_{\mathrm{rf}} \to 0$, can be obtained by calculating the time derivative of equation~(\ref{eq:7}). For this, the function 
\begin{equation}
F(I_{\mathrm{S}},t) = I_{\mathrm{S}} + I_{\mathrm{c}} \sin\left[\varphi_{\mathrm{ext}} + \varphi_{\mathrm{rf}}\sin\left(\omega t\right) + \beta_{\mathrm{L}}\frac{I_{\mathrm{S}} }{I_{\mathrm{c}}} \right] = 0
\label{eq:10}
\end{equation}
is defined to calculate the implicit derivative
\begin{eqnarray}
\diff{I_{\mathrm{S}}(t)}{t} &=& - \frac{\partial F(I_{\mathrm{S}},t) / \partial t}{\partial F(I_{\mathrm{S}},t) / \partial I_{\mathrm{S}}}  \nonumber \\ &=& - \frac{I_{\mathrm{c}}\varphi_{\mathrm{rf}} \omega \cos\left[\varphi_{\mathrm{ext}} + \varphi_{\mathrm{rf}}\sin\left(\omega t\right) + \beta_{\mathrm{L}}\frac{I_{\mathrm{S}} }{I_{\mathrm{c}}} \right]}{1 + \beta_{\mathrm{L}} \cos\left[\varphi_{\mathrm{ext}} + \varphi_{\mathrm{rf}}\sin\left(\omega t\right) + \beta_{\mathrm{L}}\frac{I_{\mathrm{S}} }{I_{\mathrm{c}}} \right]} \cos(\omega t)  \nonumber \\
&\approx& - \frac{I_{\mathrm{c}}\varphi_{\mathrm{rf}} \omega \cos\left(\varphi_{\mathrm{tot}}\right)}{1 + \beta_{\mathrm{L}} \cos\left(\varphi_{\mathrm{tot}}  \right)} \cos(\omega t).
\label{eq:11}
\end{eqnarray}
In the last transformation, $\varphi_{\mathrm{rf}} \to 0$ is assumed, yielding $\varphi_{\mathrm{tot}} \approx \varphi_{\mathrm{ext}} + \beta_{\mathrm{L}}I_{\mathrm{S}} / I_{\mathrm{c}}$. The induced current in the resonator termination can hence be calculated straightforwardly by using equation~(\ref{eq:8}):
\begin{equation}
i_{\mathrm{ind}}(t) = - \frac{I_{\mathrm{c}}\varphi_{\mathrm{rf}} M_{\mathrm{T}}}{L_{\mathrm{T}}} \frac{ \cos\left(\varphi_{\mathrm{tot}}\right)}{1 + \beta_{\mathrm{L}} \cos\left(\varphi_{\mathrm{tot}} \right)} \sin(\omega t).
\label{eq:12}
\end{equation}
It is phase-shifted by~$\pi/2$ due to the complex impedance~$i\omega L_{\mathrm{T}}$ of the resonator termination. Combining equations~(\ref{eq:9}) and (\ref{eq:12}) allows to calculate the effective resonator termination
\begin{equation}
L_{\mathrm{T,eff}}(\varphi_{\mathrm{tot}}) = L_{\mathrm{T}} - \frac{M_{\mathrm{T}}^2}{L_{\mathrm{S}}} \frac{\beta_{\mathrm{L}} \cos\left(\varphi_{\mathrm{tot}}\right)}{1 + \beta_{\mathrm{L}} \cos\left(\varphi_{\mathrm{tot}} \right)}. 
\label{eq:13}
\end{equation}
This expression is identical to equation~(\ref{eq:4}) derived within the existing $\mu$MUX model.

\subsection{\texorpdfstring{$\mu$}{µ}MUX characteristics for small SQUID hysteresis parameter\texorpdfstring{, i.e. $\beta_\mathrm{L}\rightarrow 0$}{}}
\label{sec:3.3}

For a vanishing SQUID hysteresis parameter, i.e. $\beta_{\mathrm{L}} \to 0$, the flux generated by the supercurrent within the SQUID loop becomes negligible and the third term in equation~(\ref{eq:7}) can be omitted. The supercurrent can then be decomposed into a Fourier series:

\begin{eqnarray}
I_{\mathrm{S}}(t) &\approx& - I_{\mathrm{c}} \sin \left( \varphi_{\mathrm{ext}} + \varphi_{\mathrm{rf}} \sin (\omega t) \right) \nonumber \\
&=& - I_{\mathrm{c}} \sin \left( \varphi_{\mathrm{ext}}\right) \cos\left( \varphi_{\mathrm{rf}} \sin (\omega t) \right) \nonumber  - I_{\mathrm{c}} \cos \left( \varphi_{\mathrm{ext}}\right) \sin\left( \varphi_{\mathrm{rf}} \sin (\omega t) \right) \nonumber\\
&=&- I_{\mathrm{c}} \sin\left(\varphi_{\mathrm{ext}}\right) \left[J_{0}\left(\varphi_{\mathrm{rf}}\right) + 2\sum_{i=1}^{\infty} J_{2i}\left(\varphi_{\mathrm{rf}}\right)\cos\left(2i\omega t\right)\right] \nonumber \\
&&-I_{\mathrm{c}}\cos\left(\varphi_{\mathrm{ext}}\right) \left[ 2\sum_{i=0}^{\infty} J_{2i+1}\left(\varphi_{\mathrm{rf}}\right)\sin\left([2i+1]\omega t\right)\right],
\label{eq:14}
\end{eqnarray}
where $J_{i}(x)$ denote the Bessel functions of first kind.

The supercurrent $I_\mathrm{S}(t)$ and consequently the current $i_{\mathrm{ind}}(t)$ in the resonator termination contain Fourier components with multiples of the angular frequency~$\omega$. However, due to the resonance condition (see equation~(\ref{eq:3})), only the fundamental frequency $\omega$ populates the resonator. Higher harmonics do not meet the resonance condition and hence interfere destructively within the cavity. Using equations~(\ref{eq:9}) and (\ref{eq:12}), we yield the induced current
\begin{equation}
i_{\mathrm{ind}}(t) = - \frac{2 I_{\mathrm{c}} M_{\mathrm{T}}}{ L_{\mathrm{T}}} \cos(\varphi_{\mathrm{ext}}) J_{1}(\varphi_{\mathrm{rf}}) \sin(\omega t).
\label{eq:15}
\end{equation}
within the resonator termination and consequently the flux dependence of the resonance frequency:

\begin{figure}[t]
    \centering
    \includegraphics[width=9cm]{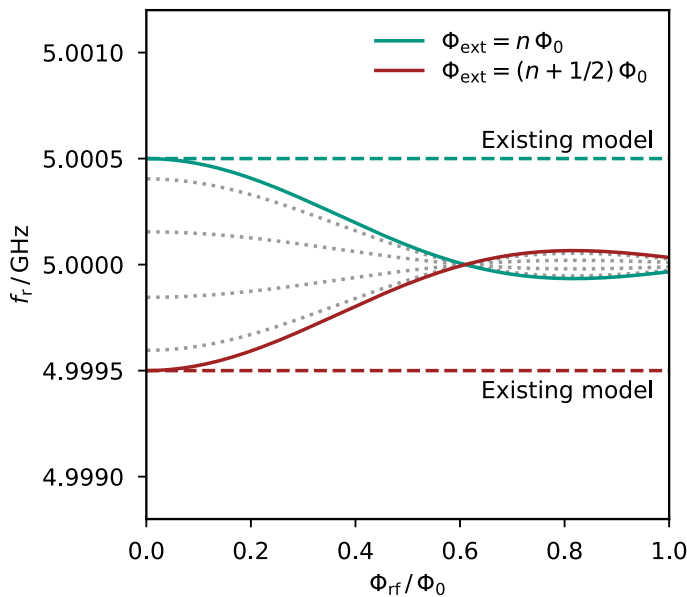}
        \caption{Dependence of the resonance frequency~$f_{\mathrm{r}}(\Phi_{\mathrm{rf}})$ on the flux amplitude~$\Phi_{\mathrm{rf}}$ of the probe signal for different values of the external flux~$\Phi_{\mathrm{ext}}$ for multiplexer with vanishing SQUID hysteresis parameter, i.e. $\beta_{\mathrm{L}} \to 0$. The curves are based on equation~(\ref{eq:16}) and their parameters are equal to the parameters chosen in figure~\ref{fig:2}. The external magnetic flux~$\Phi_{\mathrm{ext}}$ is varied between $n\Phi_0$ and $(n+1/2)\Phi_0$ in steps of $0.1\,\Phi_{0}$. For comparison, the corresponding curves of the existing model are plotted as dashed lines.}
    \label{fig:4}
\end{figure}

\begin{equation}
f_{r}(\varphi_{\mathrm{ext}}, \varphi_{\mathrm{rf}}) 
\approx f_{0}-4 f_{0}^2\left[C_{\mathrm{c}}Z_{0}+ \frac{L_{\mathrm{T}}}{Z_{0}}- \frac{M_{\mathrm{T}}^2}{Z_{0} L_{\mathrm{S}}} \frac{2 \beta_{\mathrm{L}} }{\varphi_{\mathrm{rf}}}  J_{1}(\varphi_{\mathrm{rf}})\cos(\varphi_{\mathrm{ext}})\right].
\label{eq:16}
\end{equation}
For small SQUID hysteresis parameters, i.e. $\beta_{\mathrm{L}} \to 0$, and small readout signals, i.e. $\varphi_{\mathrm{rf}} \to 0$, our model and the existing model (see equation~(\ref{eq:5})) yield identical results since $1+\beta_{\mathrm{L}} \cos(\varphi_{\mathrm{tot}}) \to 1$ and $2J_{1}(\varphi_{\mathrm{rf}})/\varphi_{\mathrm{rf}} \to 1$. However, for probing signals with finite readout power, i.e. $\varphi_{\mathrm{rf}} > 0$, the maximum resonance frequency shift $\Delta f_{\mathrm{r}}^{\mathrm{max}}$ differs from the prediction of the existing $\mu$MUX model as indicated by figure~\ref{fig:4}. With increasing readout power $\varphi_{\mathrm{rf}}$, the non-linear response of the Josephson junction converts more and more signal power into higher harmonics which do not match the resonance condition of the resonance circuit. This leads to a decrease of the maximum resonance frequency shift $\Delta f_{\mathrm{r}}^{\mathrm{max}}$ with increasing readout power. For $\Phi_{\mathrm{rf}} \approx 0.61\,\Phi_{0}$, we get $J_{1} (\varphi_{\mathrm{rf}}) = 0$  and consequently the resonance frequency $f_{\mathrm{r}}(\Phi_{\mathrm{ext}})$ gets independent of the external flux $\Phi_{\mathrm{ext}}$. With further increasing the readout power, the curves start to oscillate around the center frequency at~$f_{\mathrm{r}}(\Phi_{\mathrm{ext}} = [n \pm 1/4]\Phi_{0})$. The resonance frequencies for $\Phi_{\mathrm{ext}} = n\,\Phi_{0}$ and~$\Phi_{\mathrm{ext}} = (n+1/2)\,\Phi_{0}$ might therefore swap their position depending on the amplitude range of the probe signal.

\subsection{\texorpdfstring{$\mu$}{µ}MUX characteristics for finite hysteresis parameter and probe signal amplitude}
\label{sec:3.4}

Using trigonometric identities, we can rewrite equation~(\ref{eq:7}) as
\begin{eqnarray}
I_{\mathrm{S}} (t)
& = & - I_{\mathrm{c}} \sin\left[\varphi_{\mathrm{ext}} + \varphi_{\mathrm{rf}}\sin\left(\omega t\right) + \beta_{\mathrm{L}} \frac{I_{\mathrm{S}}(t)}{I_{\mathrm{c}}} \right]
\label{eq:17}\\
&=& - I_{\mathrm{c}} \sin \left( \varphi_{\mathrm{ext}} + \varphi_{\mathrm{rf}} \sin (\omega t)\right) \cos\left[ \beta_{L} \frac{I_{\mathrm{S}}(t)}{I_{\mathrm{c}}}  \right] \nonumber \\ 
&&- I_{\mathrm{c}} \cos \left( \varphi_{\mathrm{ext}} + \varphi_{\mathrm{rf}} \sin (\omega t)\right) \sin\left[ \beta_{L} \frac{I_{\mathrm{S}}(t)}{I_{\mathrm{c}}}  \right].
\label{eq:18}
\end{eqnarray}
Expression~(\ref{eq:18}) can be expanded as a polynomial equation by using a $2^{\mathrm{nd}}$ order Taylor expansion with respect to $\beta_{\mathrm{L}}$, i.e. by using the approximations $\cos(\beta_{\mathrm{L}} I_{\mathrm{S}}(t)/I_{\mathrm{c}}) \approx 1 - \left(\beta_{\mathrm{L}} I_{\mathrm{S}}(t)/I_{\mathrm{c}}\right)^{2}/2$ and $\sin(\beta_{\mathrm{L}} I_{\mathrm{S}}(t)/I_{\mathrm{c}}) \approx \left(\beta_{\mathrm{L}} I_{\mathrm{S}}(t)/I_{\mathrm{c}}\right)$. Solving the resulting quadratic equation for the supercurrent $I_{\mathrm{S}} (t)$ yields the solutions
\begin{eqnarray}
    I_{\mathrm{S}} (\varphi) &\approx & - I_{\mathrm{c}} \frac{-1-\beta_{\mathrm{L}} \cos(\varphi) \pm \sqrt{1+2\beta_{\mathrm{L}} \cos(\varphi) + \beta_{\mathrm{L}}^{2}\left[1+\sin(\varphi)^{2}\right]}}{\beta_{\mathrm{L}}^{2} \sin(\varphi)}
    \label{eq:19}
\end{eqnarray}
with $\varphi = \varphi_{\mathrm{ext}}+\varphi_{\mathrm{rf}}\sin(\omega t)$. Here, the solution with positive sign is relevant as the other solution leads to a divergence for $\sin(\varphi) \to 0$. 

To convert equation~(\ref{eq:19}) into a more handy form, we perform a second Taylor approximation with respect to the parameter $\beta_{\mathrm{L}}$. Within the context of this work, a $10^{\mathrm{th}}$ order Taylor expansion was used. However, higher-order terms can in general be included. Similarly to section \ref{sec:3.3}, we introduce Bessel functions $J_{1}(x)$ of first kind and neglect all terms which contain multiples of the angular frequency $\omega$ as they are not matching the resonance condition. This finally leads to
\begin{equation}
f_{\mathrm{r}}(\varphi_{\mathrm{ext}}, \varphi_{\mathrm{rf}}) 
\approx f_{0}-4 f_{0}^2\left[C_{\mathrm{c}}Z_{0}+ \frac{L_{\mathrm{T}}}{Z_{0}}- \frac{M_{\mathrm{T}}^2 }{Z_{0} L_{\mathrm{S}}} \frac{ 2\beta_{\mathrm{L}} }{ \varphi_{\mathrm{rf}}} \sum_{i,j} p_{i,j} \right]
\label{eq:20}
\end{equation}
with $p_{i,j} = a_{i,j} \beta_{\mathrm{L}}^{b_{i,j}}  J_{1}(c_{i,j} \varphi_{\mathrm{rf}}) \cos(c_{i,j}\varphi_{\mathrm{ext}})$. The parameters $a_{i,j}$, $b_{i,j}$ and $c_{i,j}$ are summarized in table~\ref{tab:1}. Here, $i$ denotes the Taylor expansion order and $j$ addresses different contributions of each order.  

\begin{table}[h]
\begin{center}
\begin{tabular}{c c} 
\begin{tabular}{|>{\centering\arraybackslash}p{8mm}|>{\centering\arraybackslash}p{26mm}|>{\centering\arraybackslash}p{5mm}|>{\centering\arraybackslash}p{5mm}|}
\hline
& $a_{i,j}$ & $b_{i,j}$ & $c_{i,j}$ \\
\hhline{|====|}
$p_{0,0}$ & $+1$ & $0$ & $1$ \\ 
\hline
$p_{1,0}$ & $-1/2$ & $1$ & $2$ \\ 
\hline
$p_{2,0}$ & $-1/8$ & $2$ & $1$ \\ 
$p_{2,1}$ & $+3/8$ & $2$ & $3$ \\ 
\hline
$p_{3,0}$ & $+1/8$ & $3$ & $2$ \\
$p_{3,1}$ & $-5/16$ & $3$ & $4$ \\ 
\hline
$p_{4,0}$ & $+1/16$ & $4$ & $1$ \\
$p_{4,1}$ & $-5/32$ & $4$ & $3$ \\
$p_{4,2}$ & $+9/32$ & $4$ & $5$ \\
\hline
$p_{5,0}$ & $-5/64$ & $5$ & $2$ \\
$p_{5,1}$ & $+3/16$ & $5$ & $4$ \\
$p_{5,2}$ & $-17/64$ & $5$ & $6$ \\
\hline
$p_{6,0}$ & $-15/512$ & $6$ & $1$ \\
$p_{6,1}$ & $+57/512$ & $6$ & $3$ \\
$p_{6,2}$ & $-115/512$ & $6$ & $5$ \\
$p_{6,3}$ & $+133/512$ & $6$ & $7$ \\
\hline
$p_{7,0}$ & $+21/512$ & $7$ & $2$ \\
$p_{7,1}$ & $-77/512$ & $7$ & $4$ \\
\hline
\end{tabular}
&
\begin{tabular}{|>{\centering\arraybackslash}p{8mm}|>{\centering\arraybackslash}p{26mm}|>{\centering\arraybackslash}p{5mm}|>{\centering\arraybackslash}p{5mm}|}
\hline
&$a_{i,j}$ & $b_{i,j}$ & $c_{i,j}$ \\
\hhline{|====|}
$p_{7,2}$ & $+137/512$ & $7$ & $6$ \\
$p_{7,3}$ & $-267/1\,024$ & $7$ & $8$ \\
\hline
$p_{8,0}$ & $+21/1\,024$ & $8$ & $1$ \\
$p_{8,1}$ & $-35/512$ & $8$ & $3$ \\
$p_{8,2}$ & $+103/512$ & $8$ & $5$ \\
$p_{8,3}$ & $-651/2\,048$ & $8$ & $7$ \\
$p_{8,4}$ & $+547/2\,048$ & $8$ & $9$ \\
\hline
$p_{9,0}$ & $-63/2\,048$ & $9$ & $2$ \\
$p_{9,1}$ & $+27/256$ & $9$ & $4$ \\
$p_{9,2}$ & $-1\,089/4\,096$ & $9$ & $6$ \\
$p_{9,3}$ & $+193/512$ & $9$ & $8$ \\
$p_{9,4}$ & $-1\,139/4\,096$ & $9$ & $10$ \\
\hline
$p_{10,0}$ & $-105/8\,192$ & $10$ & $1$ \\
$p_{10,1}$ & $+435/8\,192$ & $10$ & $3$ \\
$p_{10,2}$ & $-2\,595/16\,384$ & $10$ & $5$ \\
$p_{10,3}$ & $+5\,705/16\,384$ & $10$ & $7$ \\
$p_{10,4}$ & $-7\,317/16\,384$ & $10$ & $9$ \\
$p_{10,5}$ & $+4\,807/16\,384$ & $10$ & $11$ \\
\hline
\end{tabular}
\end{tabular}
\end{center}
\caption{Set of parameters used for calculating the resonance frequency $f_{\mathrm{r}}(\varphi_{\mathrm{ext}}, \varphi_{\mathrm{rf}})$ according to equation~(\ref{eq:20}).}
\label{tab:1}
\end{table}

\begin{figure}[t]
    \centering
    \includegraphics[width=9cm]{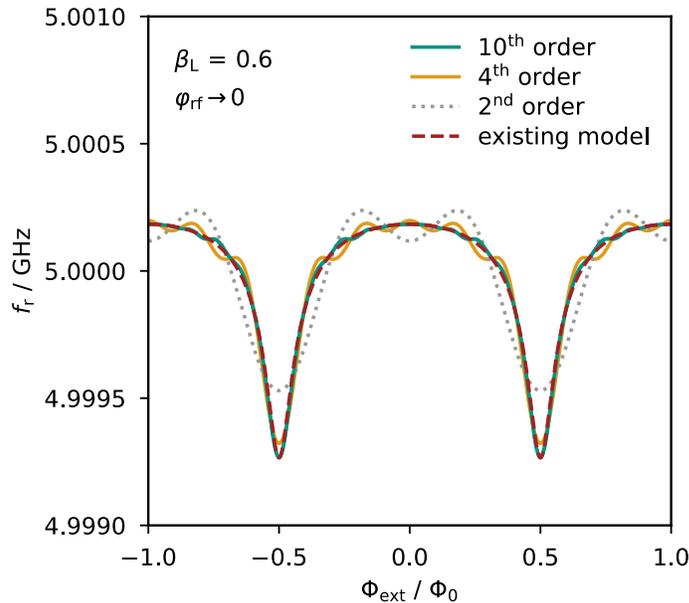}
        \caption{Flux dependence of the resonance frequency $f_{\mathrm{r}}(\Phi_{\mathrm{ext}})$ calculated by means of our $\mu$MUX model for a negligibly small amplitude of the readout signal, i.e. $\varphi_{\mathrm{rf}} \to 0$, for different values of the SQUID hysteresis parameter $\beta_{\mathrm{L}}$. The ratio $M_{\mathrm{T}}^{2}/L_{\mathrm{S}}$ is adapted to $\beta_{\mathrm{L}}$ to achieve $\Delta f_{\mathrm{r}}^{\mathrm{max}} = 1\,\mathrm{MHz}$, whereas all other parameters equal the parameters chosen in figure \ref{fig:2}.
        }
    \label{fig:5}
\end{figure}

\begin{figure}[t]
    \centering
    \includegraphics[width=9cm]{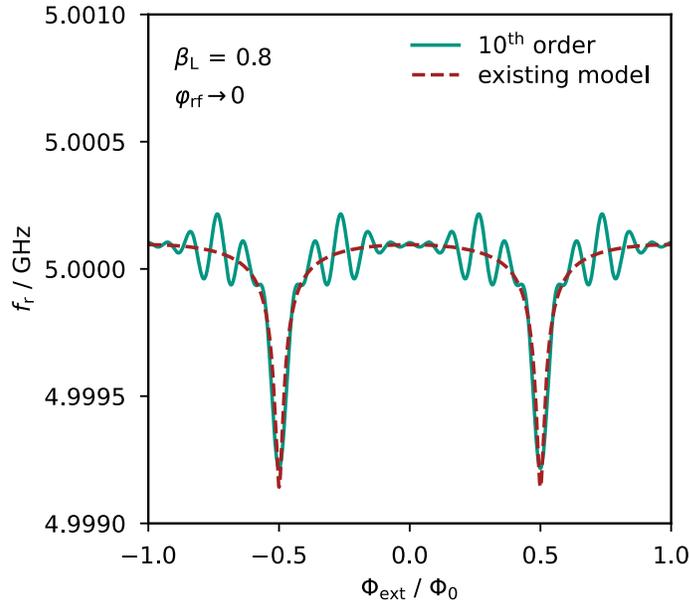}
        \caption{Flux dependence of the resonance frequency $f_{\mathrm{r}}(\Phi_{\mathrm{ext}})$ calculated by means of our $\mu$MUX model based on a $10^{\mathrm{th}}$ order Taylor expansion and the existing model for a negligibly small amplitude of the readout signal, i.e. $\varphi_{\mathrm{rf}} \to 0$, and a SQUID hysteresis parameter of $\beta_{\mathrm{L}} = 0.8$. The ratio $M_{\mathrm{T}}^{2}/L_{\mathrm{S}}$ is adapted to $\beta_{\mathrm{L}}$ to achieve $\Delta f_{\mathrm{r}}^{\mathrm{max}} = 1\,\mathrm{MHz}$, whereas all other parameters equal the parameters chosen in figure \ref{fig:2}.
        }
    \label{fig:6}
\end{figure}

To estimate the accuracy of our approximation, we show in figure~\ref{fig:5} the dependence of the resonance frequency~$f_{\mathrm{r}}(\Phi_{\mathrm{ext}})$ on the external flux~$\Phi_{\mathrm{ext}}$ for (i) the existing $\mu$MUX model and an approximation based on (ii) a $2^\mathrm{nd}$ order, (iii) a $4^\mathrm{th}$ order, and (iv) a $10^\mathrm{th}$ order Taylor expansion assuming a SQUID with $\beta_{\mathrm{L}} = 0.6$ and a probing signal with negligible readout power, i.e. $\varphi_\mathrm{rf} \to 0$. The curve obtained by means of the existing model is exact for the chosen parameter range, i.e. $\varphi_\mathrm{rf} \to 0$, and
essentially equals the corresponding curve shown in figure~\ref{fig:3}. However, in contrast to figure~\ref{fig:3}, a transformation~$\Phi_{\mathrm{tot}} \to \Phi_{\mathrm{ext}}$ was applied for allowing a direct comparison between the different models. The plot demonstrates a large deviation between the exact model and approximations based on low-order Taylor expansions and confirms a very good agreement between the existing model and the approximation based on a $10^\mathrm{th}$ order Taylor expansion. Therefore, lower-order approximations are only valid for small hysteresis parameters~$\beta_{\mathrm{L}}$. In the same way, the approximation based on a $10^\mathrm{th}$ order Taylor expansion becomes more and more imprecise for $\beta_{\mathrm{L}} > 0.6$. For example, for $\beta_{\mathrm{L}} = 0.8$ the characteristics based on the $10^\mathrm{th}$ order model deviate significantly from the exact model (see figure \ref{fig:6}). To conclude, these results indicate that our model is valid in the range of $ \beta_{\mathrm{L}} \le 0.6$ and can potentially be fitted to $ \beta_{\mathrm{L}} > 0.6$ when using higher-order Taylor expansions. However, for existing $\mu$MUXs, the hysteresis parameter is typically $ \beta_{\mathrm{L}} < 0.6$, thus our model based on the $10^{\mathrm{th}}$ order Taylor expansion covers the relevant parameter range.

\begin{figure}[t]
    \centering
    \includegraphics[width=9cm]{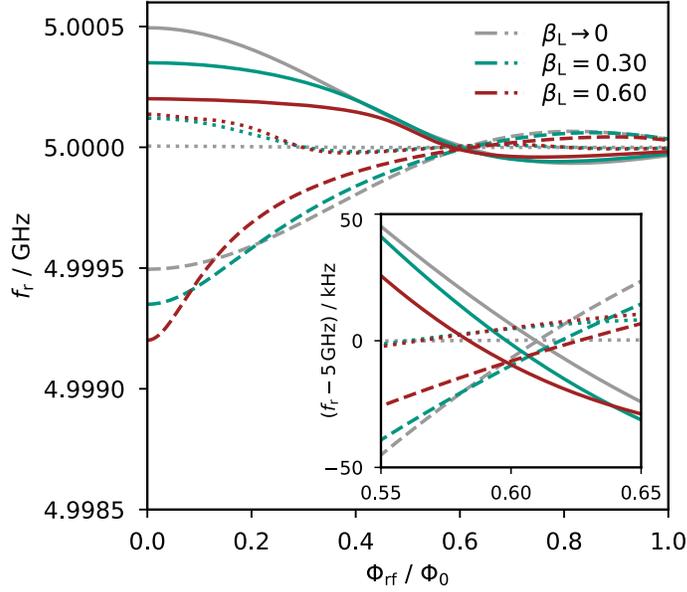}
    \caption{Dependence of the resonance frequency~$f_{\mathrm{r}}(\Phi_{\mathrm{rf}})$ on the flux amplitude~$\Phi_{\mathrm{rf}}$ of the readout signal calculated by using our model being based on a $10^\mathrm{th}$ order Taylor expansion for different SQUID hysteresis parameters~$\beta_{\mathrm{L}}$. The values of the magnetic flux are  $\Phi_{\mathrm{ext}} = n\,\Phi_0$ (solid lines), $\Phi_{\mathrm{ext}} = (n\pm1/4)\Phi_0$  (dotted lines) and $\Phi_{\mathrm{ext}} = (n+1/2)\Phi_0$ (dashed lines). The ratio~$M_{\mathrm{T}}^{2}/L_{\mathrm{S}}$ for all curves is chosen such that the maximum resonance frequency shift~$\Delta f_{\mathrm{r}}^{\mathrm{max}} (\varphi_{\mathrm{rf}} \to 0) = 1\,\mathrm{MHz}$ is identical for different values of~$\beta_{\mathrm{L}}$, whereas all other curve parameters correspond to the parameters chosen in figure~\ref{fig:2}.}
    \label{fig:7}
\end{figure}

Figure \ref{fig:7} shows the dependence of the resonance frequency $f_{\mathrm{r}}(\Phi_{\mathrm{rf}})$ on the amplitude $\Phi_{\mathrm{rf}}$ of the probing signal for different values of the external flux $\Phi_{\mathrm{ext}}$ and the hysteresis parameter $\beta_{\mathrm{L}}$ as calculated using our model. The characteristics for $\beta_{\mathrm{L}} \to 0$ describe the limit of vanishing self-induced flux ($\Phi_{\mathrm{scr}} \to 0$), and therefore correspond to the case shown in figure \ref{fig:4}. The curves for $\beta_{\mathrm{L}} > 0$ follow a similar course, but show asymmetric shapes due to non-linearities resulting from a finite flux contribution $\Phi_{\mathrm{scr}} > 0$ of the screening current. This effect becomes particularly important for $\Phi_{\mathrm{rf}} \to 0$, while the influence of $\beta_{\mathrm{L}}$ on the characteristics gets smaller when increasing the readout power. It should be noted that the curves do not intersect exactly at the first zero of the Bessel function $J_{1}(\varphi_{\mathrm{rf}})$, as it is the case for $\beta_{\mathrm{L}} \to 0$. However, as one can see in the inset graph, this effect is only very small.

\begin{figure}[t]
    \centering
    \includegraphics[width=9cm]{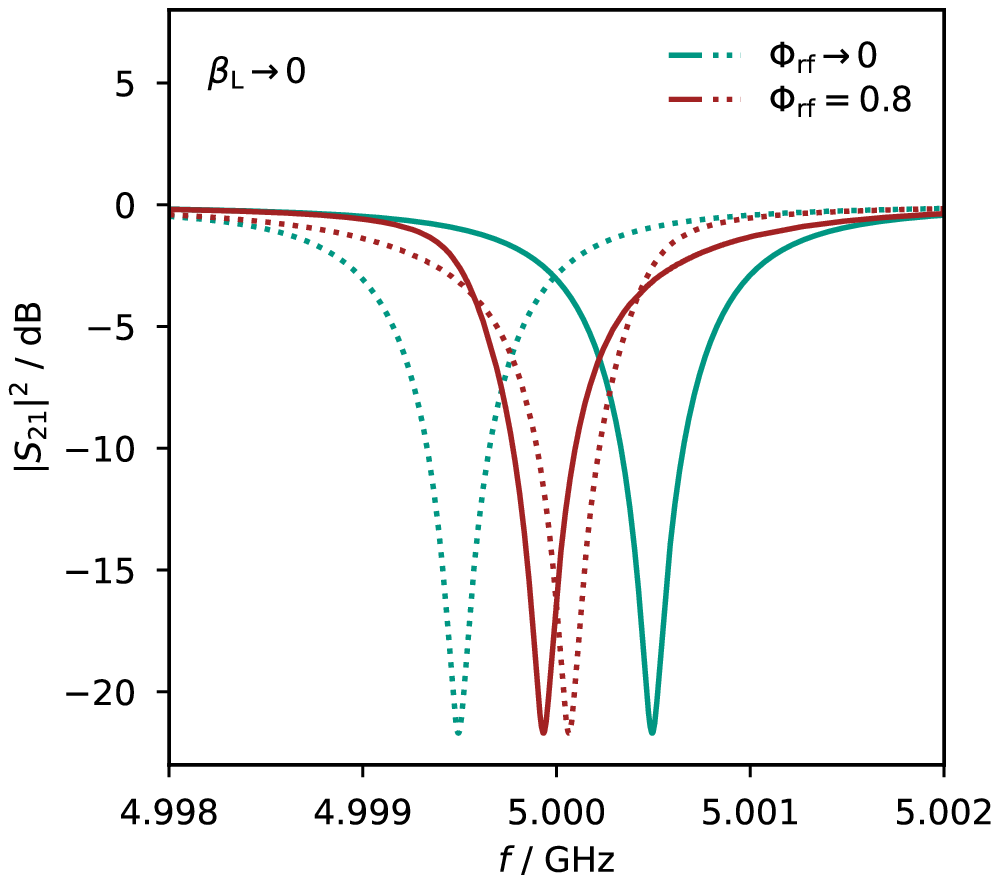}
    \caption{Frequency dependence of the transmission amplitude $|S_{21}(f)|^{2}$ in the low power limit and for high readout powers. The values of the magnetic flux are  $\Phi_{\mathrm{ext}} = n\,\Phi_0$ (solid lines) and $\Phi_{\mathrm{ext}} = (n+1/2)\Phi_0$ (dotted lines).  All curves are numerically calculated, using our model and an iterative method. While the characteristics for very low readout power are identical to the curves shown in figure \ref{fig:2}, the curves for high readout power are strongly asymmetric.}
    \label{fig:8}
\end{figure}

Figure~\ref{fig:8} shows another non-linear effect which is predicted by our model. Here, the resonance curves with minimum and maximum resonance frequency, i.e. for $\Phi_{\mathrm{ext}} = n\,\Phi_{0}$ and $\Phi_{\mathrm{ext}} = (n+1/2)\,\Phi_{0}$, are plotted for $\Phi_{\mathrm{rf}} \to 0$ and $\Phi_{\mathrm{rf}} = 0.8\,\Phi_{0}$. As one can see,  the maximum resonance frequency shift is much smaller for high readout powers, and the resonance curves swap their position, i.e. $f_{\mathrm{r}}(n\,\Phi_{0}) < f_{\mathrm{r}}([n+1/2]\,\Phi_{0})$. These effects are expected and in good agreement with the results presented in figure~\ref{fig:7}. However, the resonance curves do not get shifted as a whole, but rather become distorted and asymmetric. This asymmetry is a result of the frequency-dependent amplitude of the driving current in the resonator termination. The maximum amplitude $I_{\mathrm{T}}$ (see equation~(\ref{eq:6})) is only reached exactly on resonance. Hence, for $f \neq f_{\mathrm{r}}$, the power in the resonator and therefore $\Phi_{\mathrm{rf}}$ are always smaller than the maximum value. As a consequence, the power-dependent resonance shift is at maximum on resonance, whereas for $f \neq f_{\mathrm{r}}$, the red lines for $\Phi_{\mathrm{rf}} = 0.8\,\Phi_{0}$ tend to approach the corresponding green lines for $\Phi_{\mathrm{rf}} \to  0$ instead of keeping their shape. Analytical expressions which describe these typical resonance curve asymmetries could not be derived within the context of this work, which is why the characteristics shown in figure \ref{fig:8} are calculated numerically, using an iterative method in combination with our analytical approximation model.

\section{Comparison between our model and measured \texorpdfstring{$\mu$}{µ}MUX characteristics}

For verification of our $\mu$MUX model, data from measurements described in~\cite{kem17} were used. Here, the scattering parameter~$S_{21}(f, \Phi_{\mathrm{ext}}, \Phi_{\mathrm{rf}})$ of various channels of a $\mu$MUX device for different values of the external fluxe~$\Phi_{\mathrm{ext}}$ and the amplitude $\Phi_{\mathrm{rf}}$ of the driving signal were measured using a vector network analyzer~(VNA). While the external flux $\Phi_{\mathrm{ext}}$ was varied by changing the current $I_{\mathrm{mod}}$ through the modulation coil of the $\mu$MUX normally used for flux ramp modulation, the amplitude $\Phi_{\mathrm{rf}}$ of the probing signal was varied by changing the VNA signal power $P_{\mathrm{VNA}}$ and hence the resonator readout power~$P_{\mathrm{rf}}$. For extracting the relevant resonator parameters from measured raw data, we applied the algorithm described in~\cite{pro15}.

\begin{figure}[ht]
    \centering
    \includegraphics[width=9cm]{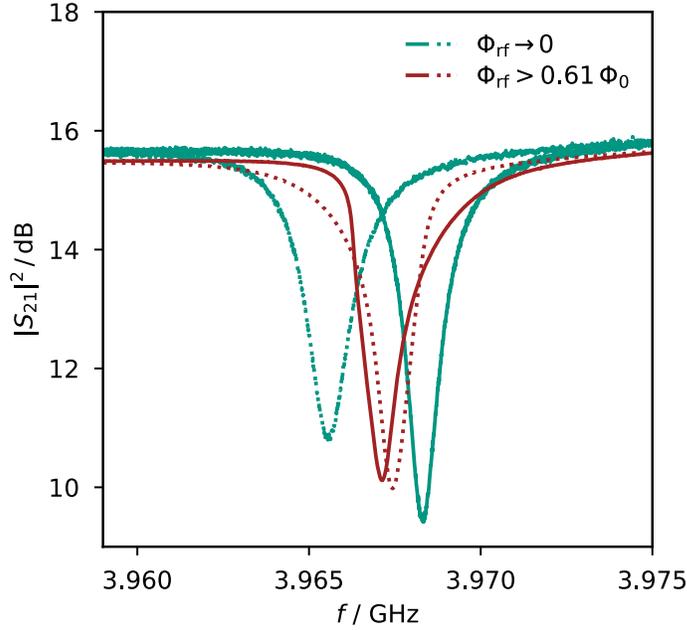}
    \caption{Measured transmission amplitude $|S_{21}(f)|^{2}$ for two different high frequency flux amplitudes $\Phi_{\mathrm{rf}}$. The values of the magnetic flux are  $\Phi_{\mathrm{ext}} = n\,\Phi_0$ (solid lines) and $\Phi_{\mathrm{ext}} = (n+1/2)\Phi_0$ (dotted lines). While the resonance curves for a negligibly small amplitude $\Phi_{\mathrm{rf}}$ have a symmetric shape, the curves recorded at high resonator readout power $P_{\mathrm{rf}}$ show large asymmetries. In addition, the resonance frequencies of both flux states~$\Phi_{\mathrm{ext}}$ start to switch their position as expected by the prediction of the power-dependent $\mu$MUX model.}
    \label{fig:9}
\end{figure}

Figure~\ref{fig:9} shows the transmission amplitude $|S_{21}(f)|^{2}$ of an arbitraty $\mu$MUX channel for $\Phi_{\mathrm{ext}} = n\,\Phi_{0}$ and $\Phi_{\mathrm{ext}} = (n+1/2)\,\Phi_{0}$. The measurement indicated by green color was performed at small readout power, i.e. $P_{\mathrm{rf}} \to 0$. As expected by our model, both resonance curves have a symmetric shape and the resonance frequency is at maximum for $\Phi_{\mathrm{ext}} = n\,\Phi_{0}$ and at minimum for $\Phi_{\mathrm{ext}} = (n+1/2)\,\Phi_{0}$. In contrast, the resonance curves recorded at $P_\mathrm{VNA} = -15\,\mathrm{dBm}$ switched their position, i.e. $f_{\mathrm{r}}(\Phi_{\mathrm{ext}} = n\,\Phi_{0}) < f_{\mathrm{r}}(\Phi_{\mathrm{ext}} = [n+1/2]\,\Phi_{0})$, and become asymmetric, both in full agreement with our model. However, in contrast to our model, the curves show different depths which depend on the actual value $\Phi_{\mathrm{ext}}$ of the external flux. This results from the magnetic flux dependence of the internal quality factor $Q_{\mathrm{i}}(\Phi_{\mathrm{ext}})$ of the $\mu$MUX channel as discussed in detail in \cite{kem17}. This effect isn't included in the model presented here as we are focusing solely on the dependence of the resonance frequency on readout power and SQUID hysteresis parameter.
\begin{figure}[t]
    \centering
    \includegraphics[width=9cm]{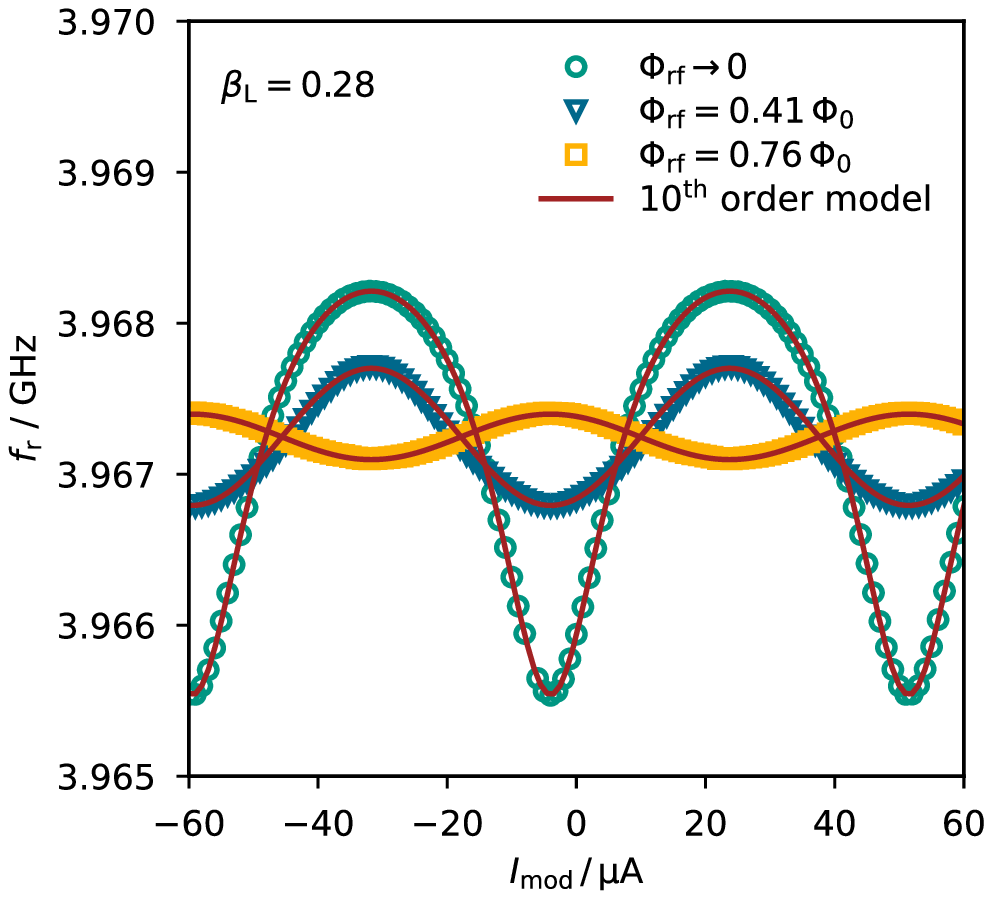}
    \caption{Measured resonance frequency $f_{\mathrm{r}}(I_{\mathrm{mod}})$ as a function of the modulation coil current $I_{\mathrm{mod}}$ for different high frequency flux amplitudes $\Phi_{\mathrm{rf}}$. In addition, for each curve a numerical fit is shown which is based on equation (\ref{eq:21}). While one curve was recorded at very low readout powers, i.e. $\Phi_{\mathrm{rf}} \to 0$, the determined high frequency flux amplitude $\Phi_{\mathrm{rf}}$ for the other curves is a result of the numerical fits.}
    \label{fig:10}
\end{figure}

For analyzing the dependence of the resonance frequency on the applied external flux $\Phi_{\mathrm{ext}}$, a current $I_{\mathrm{mod}}$ was sent through the common modulation coil of the $\mu$MUX, generating a flux contribution of $\Phi_{\mathrm{ext}} = M_{\mathrm{mod}} I_{\mathrm{mod}} + \Phi_{\mathrm{off}}$ threading the SQUID loop. Here, $M_{\mathrm{mod}}$ denotes the mutual inductance between modulation coil and rf-SQUID and $\Phi_{\mathrm{off}}$ a random, constant flux offset in the SQUID loop. The injected current was varied from $I_{\mathrm{mod}} = -60\,\mathrm{\mu A}$ to $I_{\mathrm{mod}} = 60\,\mathrm{\mu A}$ in steps of $1\,\mathrm{\mu A}$ for different readout powers $P_{\mathrm{rf}}$, leading to the three characteristics shown in figure~\ref{fig:10}. For the curve with lowest readout power, i.e. $P_{\mathrm{rf}} \to 0$, the parameter $\Phi_{\mathrm{rf}} \to 0$ was set to a fixed value. Afterwards, a numerical fit motivated by equation (\ref{eq:20}) and based on the expression
\begin{equation}
    f_{\mathrm{r}}(I_{\mathrm{mod}}, \Phi_{\mathrm{rf}}) = f_{\mathrm{r,off}} + \Delta f_{\mathrm{r,mod}} \frac{2\beta_{\mathrm{L}}}{\varphi_{\mathrm{rf}}}\sum_{i} p_{i} (I_{\mathrm{mod}}, \Phi_{\mathrm{rf}})
    \label{eq:21}
\end{equation}
was performed. Subsequently, after determining and fixing the values of the parameters $f_{\mathrm{r,off}}$, $\Delta f_{\mathrm{r,mod}}$, $\beta_{\mathrm{L}}$, $ M_{\mathrm{mod}}$ and $\Phi_{\mathrm{off}}$, the same fitting procedure was applied to the other curves where the only free fitting parameter was $\Phi_{\mathrm{rf}}$. As expected by our model, the maximum resonance frequency shift $\Delta f_{\mathrm{r}}^{\mathrm{max}}$ is largest for $\Phi_{\mathrm{rf}} \to 0$ and decreases for higher applied readout signal amplitudes. Additionally, a sign change of the modulation can be observed solely for the curve with a high frequency flux amplitude of $\Phi_{\mathrm{rf}} = 0.76\,\Phi_{0}$, which is also in good agreement with our multiplexer model. 

\begin{figure}[t]
    \centering
    \includegraphics[width=9cm]{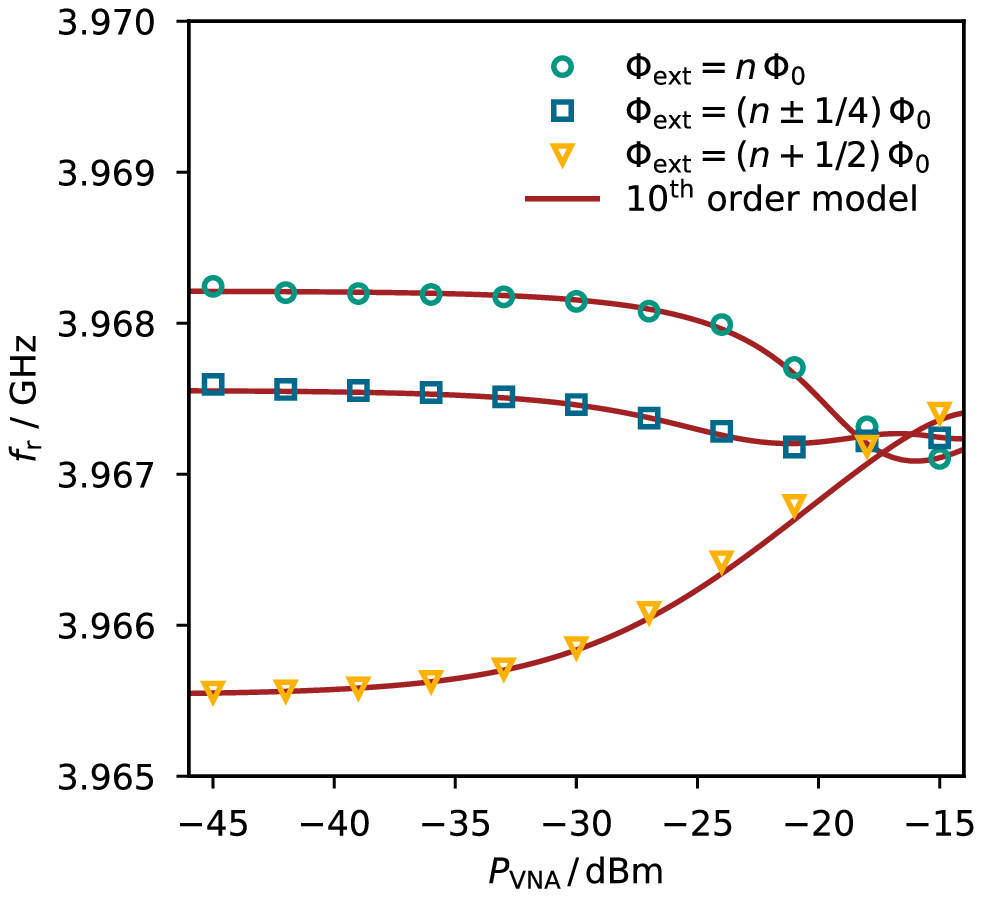}
    \caption{Measured resonance frequency $f_{\mathrm{r}}(P_{\mathrm{VNA}})$ as a function of the VNA signal power $P_{\mathrm{VNA}}$ for different values of the external flux $\Phi_{\mathrm{ext}}$. The numerical fits are based on equation (\ref{eq:21}) and contain only one free parameter which is the attenuation factor $A_{\mathrm{rf}}$, that considers the total attenuation of the high frequency components between the VNA sending port and the $\mu$MUX chip. All other curve parameters are set to fixed values.}
    \label{fig:11}
\end{figure}

Figure \ref{fig:11} shows the dependence of the measured resonance frequency $f_{\mathrm{r}}(P_{\mathrm{VNA}})$ on the signal power $P_{\mathrm{VNA}}$ for different values of the external flux $\Phi_{\mathrm{ext}}$. Here, the modulation current $I_{\mathrm{mod}}$ was set to a fixed value to achieve the corresponding external flux $\Phi_{\mathrm{ext}}$, and the VNA readout power was varied from $P_{\mathrm{VNA}} = -45\,\mathrm{dBm}$ to $P_{\mathrm{VNA}} = -15\,\mathrm{dBm}$ in steps of $3\,\mathrm{dBm}$. To convert VNA signal power $P_{\mathrm{VNA}}$ into readout power $P_{\mathrm{rf}}$, an attenuation factor $A_{\mathrm{rf}} < 1$ of the cryogenic microwave setup was introduced which is assumed to be independent of power and frequency, i.e. $P_{\mathrm{rf}} = A_{\mathrm{rf}} P_{\mathrm{VNA}}$. This attenuation factor represents the attenuation of coaxial cables and high frequency components which are placed in between the VNA sending port and the $\mu$MUX chip on the cold stage of the cryostat. It is the only free parameter of the numerical fits, which are shown in the graph as well and which are based on equations (\ref{eq:6}) and (\ref{eq:21}). While the quality factors $Q_{\mathrm{l}}$ and $Q_{\mathrm{c}}$ in equation (\ref{eq:6}) are determined and fixed by means of the applied resonance curve analysis algorithm described in \cite{pro15}, all other $\mu$MUX parameters are set to the values given by the numerical fits shown in figure \ref{fig:10}. At this point, the flux dependence of the internal quality factor $Q_{\mathrm{i}}(\Phi_{\mathrm{ext}})$ observed in figure~\ref{fig:9} must be considered. As a result of different quality factors, the current $I_{\mathrm{T}}(\Phi_{\mathrm{ext}})$ in the resonator termination and therefore the generated high frequency flux amplitude $\Phi_{\mathrm{rf}}(\Phi_{\mathrm{ext}})$ are flux dependent as well. As a consequence, the numerical fit for $\Phi_{\mathrm{ext}} = n\,\Phi_{0}$ with higher internal quality factor $Q_{\mathrm{i}}$ shifts towards lower VNA signal powers $P_{\mathrm{VNA}}$, whereas the curve with $\Phi_{\mathrm{ext}} = (n+1/2)\,\Phi_{0}$ moves towards higher VNA signal powers $P_{\mathrm{VNA}}$. As a result, the intersections of the three numerical fits in figure~\ref{fig:11} are not as close together as shown in figure~\ref{fig:7}. Considering this effect, we yield a very good agreement between our model and experimental data.

\section{Conclusion}

We presented an analytical model being able to describe the readout power dependence of the resonance frequency on the amplitude $P_{\mathrm{rf}}$ of the readout signal as well as the SQUID hysteresis parameter $\beta_{\mathrm{L}}$. For this, we analyzed the Fourier components of the non-linear response of the non-hysteretic rf-SQUID. We were able to to derive an analytical approximation describing the multiplexer in the parameter range of $\beta_{\mathrm{L}} < 0.6$ and showed that our model includes the existing model as a limiting case. We verified our model by comparing it to measurements. Here, we demonstrated a very good agreement between measured multiplexer characteristics and model predictions. Our model hence allows for a deep understanding of the complex $\mu$MUX behavior, a prerequisite for a full device optimization as being required for realizing next-generation detector instruments strongly relying on the use of SQUID based multiplexing techniques with high multiplexing factor. 

\section{Acknowledgments}
This work was performed within the framework of the DFG research unit FOR 2202 (funding under grant no. EN299/7-1, EN299/7-2, EN299/8-1, and EN299/8-2). The research leading to these results has also received funding from the European Union’s Horizon 2020 Research and Innovation Programme, under Grant Agreement No. 824109.

\section{References}
\bibliographystyle{iopart-num}
\bibliography{bib.bib}

\end{document}